\documentclass[journal]{IEEEtran}
\usepackage{cite}
\usepackage{amsmath,amssymb,amsfonts}
\usepackage{algorithmic}
\usepackage{graphicx}
\usepackage{textcomp}
\usepackage{xcolor}
\usepackage{subfig}
\usepackage{booktabs}
\usepackage{import}
\usepackage{color}
\usepackage[citebordercolor=blue]{hyperref}
\usepackage[capitalise]{cleveref}
\usepackage{multirow}
\usepackage{soul}
\usepackage{array}
\usepackage[font=footnotesize]{caption}
\usepackage[inline]{enumitem}
\usepackage{makecell}
\usepackage{siunitx}
\usepackage{caption}
\usepackage{comment}

\begin{document}
\bstctlcite{BSTControl}
%

\title{Digital Twin Satellite Networks: A Paradigm for Intelligent, Efficient, and Resilient Operations}

%
%
%

\author{Mustafa~Alhassan~\IEEEmembership{Member,~IEEE,}~and Peng~Hu~\IEEEmembership{Senior Member,~IEEE}
\thanks{Mustafa Alhassan and Peng Hu are with the Dept. of Electrical and Computer Engineering, University of Manitoba, Winnipeg, Canada. Corresponding author: Peng Hu (peng.hu@umanitoba.ca)}
}

%
%

\markboth{Journal of \LaTeX\ Class Files,~Vol.~00, No.~0, 00~2026}%
{Shell \MakeLowercase{\textit{et al.}}: Bare Demo of IEEEtran.cls for IEEE Journals}
%

\maketitle

\begin{abstract}
Satellite mega-constellations in Low Earth Orbit (LEO) are becoming an important part of next-generation non-terrestrial networks, but their operation remains challenging because of fast network topology variation, intermittent inter-satellite links, hardware disturbances, and strict Size, Weight, and Power (SWaP) constraints. Existing approaches based on Digital Twin (DT), Digital Twin Network (DTN), Software-Defined Networking (SDN), and Open Radio Access Network (O-RAN) provide useful building blocks for intelligent satellite networking, but they do not fully support real-time, predictive, and platform-aware network operation. In this paper, we propose a Digital Twin Satellite Network (DTSN) framework as a closed-loop architecture for reliable and intelligent management of LEO satellite constellations. The proposed framework connects the physical satellite network with a synchronized virtual twin and combines real-time telemetry, Integrated Sensing and Communication (ISAC), predictive intelligence, and resilience-oriented control. To validate the concept, we develop a constellation-scale cross-domain co-simulation using the NASA 42 spacecraft simulator and a Python-based DT bridge for a LEO constellation. The DT continuously ingests physical telemetry to manage a multi-domain threat environment, encompassing kinematic drift, hardware failures, and adversarial jamming over a 600-second flight window. By leveraging a predictive lookahead mechanism and an exponential sensor recovery model, the framework successfully isolates compromised nodes and triggers proactive network reconfiguration, thereby ensuring uninterrupted service and dynamic network resilience. These results show the potential of DTSN to support predictive and resilience-oriented satellite network operations.
\end{abstract}

\begin{IEEEkeywords}
Satellites, digital twin, satellite network operations, intelligent operations
\end{IEEEkeywords}

%
\IEEEpeerreviewmaketitle

\section{Introduction}
%
%
%
%

Satellite mega-constellations in Low Earth Orbit (LEO) are becoming an important part of next-generation networks for providing seamless global connectivity. However, operating these systems remains challenging because they function in highly dynamic space environments, where frequent topology changes and unexpected hardware disturbances are common. At the same time, LEO satellites are subject to strict Size, Weight, and Power (SWaP) constraints, which significantly limit their onboard computing capability and energy availability. Under these conditions, traditional reactive network management approaches, which respond only after link failure or service degradation has already occurred, are often insufficient to maintain reliable constellation operation.

To address these challenges, the Digital Twin (DT) paradigm has emerged as a promising approach. A DT is a high-fidelity virtual replica of a real physical system or process. Through continuous synchronization with its physical counterpart, it can support real-time monitoring, performance optimization, predictive maintenance, and closed-loop control. At the network level, related paradigms such as Digital Twin Networks (DTNs), Software-Defined Networking (SDN), and Open Radio Access Network (O-RAN) provide important capabilities for programmability and intelligent control. However, in satellite applications, these approaches still face important limitations. In particular, they often suffer from synchronization delays and mainly focus on network-level performance metrics, while giving limited attention to the physical health of the satellite platform, such as power status, thermal variation, and hardware behavior. Therefore, a unified framework that links network control decisions with real-time physical spacecraft conditions is needed to support proactive and reliable satellite network management.

In this paper, we first review the main research streams related to intelligent satellite network operation and identify the remaining gap in real-time, predictive, and hardware-aware satellite management. We then propose the Digital Twin Satellite Network (DTSN) framework and evaluate its operational principles through a cross-domain co-simulation study. In this context, cross-domain refers to the joint treatment of physical spacecraft dynamics, hardware-layer component states, and security-layer disturbances within a unified experimental framework. Co-simulation refers to the coordinated execution of these domains, seamlessly synchronizing the continuous orbital-physics model generated by NASA 42 with the discrete network logic of the Python-based digital twin along a shared timeline. Unlike conventional static simulation approaches, the DTSN is designed as a real-time, bidirectional control architecture that continuously interacts with the physical network. To reduce the computational load on resource-constrained satellites, the framework offloads intensive processing tasks to high-performance virtual replicas hosted on ground station servers. In addition, the DTSN incorporates the Integrated Sensing and Communication (ISAC) paradigm over highly directional Optical Inter-Satellite Links (O-ISLs), enabling the system to relate physical spacecraft behavior to communication link quality in real time.

The main contributions of this paper are summarized as follows. First, we propose a four-layer DTSN architecture that logically separates computationally intensive intelligence and decision-making functions from resource-constrained physical satellite hardware. Second, we apply the ISAC principle by translating physical attitude behavior into predictive communication metrics associated with Signal-to-Noise Ratio (SNR) degradation. Third, we introduce a predictive intelligence mechanism that supports early assessment of kinematic, hardware, and security-related disruptions before they propagate into wider network degradation. Finally, we validate the dynamic resilience of the DTSN through a closed-loop cross-domain co-simulation over a 60-node Walker Delta constellation, showing that the framework can isolate and manage overlapping multi-domain disturbances in real time.

The remainder of this paper is organized as follows. Section II reviews related work on DT, DTN, SDN-based satellite networking, and O-RAN-based satellite operations. Section III presents the proposed DTSN architecture and its core operational features. Section IV describes the cross-domain co-simulation framework, baseline constellation topology, cross-domain threat emulation, and quantitative performance results. Section V discusses perspectives on future directions, including the potential integration of quantum sensing for advanced fault diagnosis.

\section{Related Work}

As previously discussed, a DT is a digital representation of a real-world system. In networking, the virtual model is built from the observed behavior of the real system, and the insights obtained in the digital space can then guide actions in the physical system.

\subsection{Digital Twin}
In recent years, DT has evolved from mainly static physics-based simulation toward dynamic and data-driven replicas of physical systems. Today, these models can continuously synchronize with their physical counterparts to support monitoring and operational control \cite{guo_five_2024}.  While initial iterations relied on static modeling and offline analysis, modern DTs utilize big data analytics and artificial intelligent (AI) or machine learning to simulate system behavior in real-time, encompassing the full life cycle of the physical asset \cite{masaracchia_digital_2022}\cite{poorzare_network_2025}. In the context of sixth-generation (6G) wireless networks, DTs have emerged as an important tool for enabling autonomous and adaptive system operation \cite{li_digital_2025}\cite{guo_five_2024}. Today, these frameworks are widely used in the Industrial Internet of Things (IIoT), allowing researchers to run highly accurate simulations and mirror the real-time behavior of a system's control, communication, and computing (3C) \cite{xu_survey_2023}.

Network research is gradually moving from slow, resource-intensive simulation tools to more proactive and predictive modeling approaches. While discrete-event simulators such as ns-3 and OMNeT++ have been essential for protocol testing, their high computational cost and lack of real-time prediction limit their usefulness in highly dynamic environments \cite{tran_network_2025}\cite{poorzare_network_2025}. Most existing analytical methods are still reactive, meaning they address network problems only after performance has already degraded in the real physical system \cite{chen_predictive_2026}. In contrast, emerging predictive models rely on analytical performance limits to anticipate potential risks in operations prior to performance degradation. This transition requires simulation tools to run faster than real-time to test "what-if" scenarios and proactively adjust system parameters based on predicted behavior \cite{chen_predictive_2026}.
Despite the widespread use of AI, a critical gap remains: most systems lack true bidirectional synchronization. Recent reviews show that much of the current research still evaluates AI components using historical or synthetic data, instead of enabling real-time interaction between the physical system and its virtual replica, which is a core requirement of a true DT \cite{kreuzer_artificial_2024}.

\subsection{Digital Twin Network (DTN)}
While the DT framework was originally established to model discrete physical assets, its scope has expanded significantly to address the complexities of modern connectivity. This transition is marked by the development of DTNs, which adapt foundational DT principles to the systemic orchestration of entire telecommunication infrastructures \cite{yang_systematic_2021-1}\cite{tu_research_2023}. A DTN serves as a virtual representation of a physical network that continuously mirrors real network states. Through this real-time interaction, operators can monitor system behavior, diagnose performance issues, evaluate potential actions through simulation, and apply control decisions within a closed-loop framework \cite{zhou_hierarchical_2023}\cite{zhu_knowledge_2021}. This paradigm moves network management away from static, topology-based models toward data-driven routing and proactive optimization, where both historical records and real-time traffic measurements guide operational decisions \cite{wei_data-driven_2021}. At the core of this evolution is the coordinated integration of four fundamental components: data, models, mapping mechanisms, and operational interfaces. Together, these elements enable continuous support across the entire network life cycle, from accurate system planning and streamlined deployment to intelligent operation and maintenance \cite{wei_data-driven_2021}\cite{chen_data_2023}. By using high-fidelity virtual replicas of real networks, DTNs make it possible to test new technologies and service configurations in a controlled environment before deploying them in practice. This approach helps reduce operational risks and minimizes the cost and uncertainty associated with trial-and-error experimentation \cite{wang_elastic_2023}\cite{chen_data_2023}.
Recently, DTN frameworks have been extended to Space-Air-Ground Integrated Networks (SAGIN) and non-terrestrial networks (NTN) to address the growing demand for seamless global connectivity envisioned for the 6G era \cite{duong_machine_2023}\cite{gao_plotinus_nodate}. In contrast to terrestrial DTNs, satellite-based DT systems must handle the highly dynamic nature of LEO constellations, where satellites move rapidly and inter-satellite links (ISLs) change frequently over time \cite{mao_digital_2024}\cite{duong_machine_2023}. Recent studies have proposed hierarchical DT architectures in which local twins are placed at distributed ground stations to support low-latency, real-time operations, while centralized twins at network control centers handle global coordination, analysis, and long-term optimization \cite{zhou_hierarchical_2023}. In addition, DTs are being introduced to improve the security of satellite internet systems by offering a controlled environment where potential threats such as link-flooding attacks, satellite hijacking, and eavesdropping can be safely studied and evaluated \cite{lai_space_2024}. Advanced emulation platforms such as Plotinus use microservice-based architectures to reproduce interactions across physical and network layers, while integrating real network traffic to improve path computation and channel capacity modeling \cite{gao_plotinus_nodate}.

Although existing DTN approaches show strong potential, the literature indicates that the trade-off between model accuracy and synchronization delay is still an open challenge. Hierarchical designs try to reduce latency by deploying DTs closer to the network edge, but the long propagation distances in satellite systems continue to limit real-time, two-way data exchange between physical networks and their virtual counterparts. In addition, the limited computing power and storage available on satellites make it difficult to support the large data collection and intensive processing needed for high-fidelity DT models \cite{zhou_hierarchical_2023}\cite{lai_space_2024}. Many existing frameworks use simplified assumptions or lightweight models to remain scalable, but this often reduces the reliability of their results when applied to the highly dynamic behavior of large satellite constellations \cite{mao_digital_2024}. A major challenge in future DTN frameworks is maintaining accurate and real-time synchronization across thousands of fast-moving satellites while operating with limited onboard resources. This problem makes it difficult to ensure that the DT accurately reflects the state of the physical network. Addressing this challenge will likely require more advanced predictive methods as well as more efficient approaches for integrating asynchronous data from distributed satellite nodes.

\begin{table*}[t]
\caption{Summary of related work and motivation toward the proposed DTSN framework}
\label{tab:related_work_summary}
\centering
\renewcommand{\arraystretch}{1.18}
\setlength{\tabcolsep}{4pt}
\footnotesize
\begin{tabular}{|p{2.2cm}|p{3.2cm}|p{3.2cm}|p{3.8cm}|p{3.8cm}|}
\hline
\textbf{Research Stream} & \textbf{Main Focus} & \textbf{Key Strengths} & \textbf{Main Limitations in Satellite Context} & \textbf{Motivation for DTSN} \\
\hline

Digital Twin (DT) \cite{guo_five_2024}\cite{masaracchia_digital_2022}\cite{poorzare_network_2025}\cite{li_digital_2025}
\cite{xu_survey_2023}\cite{tran_network_2025}\cite{chen_predictive_2026}\cite{kreuzer_artificial_2024}
& Virtual replica of a physical system with data-driven synchronization for monitoring, simulation, and control. 
& Enables real-time system mirroring, predictive analysis, what-if evaluation, and life-cycle management. 
& Many existing DT studies remain AI-assisted but not fully bidirectional; real-time interaction between physical and virtual entities is often limited. General DT literature is also not tailored to dynamic satellite network operations. 
& Motivates a satellite-specific twin framework with continuous synchronization, closed-loop operation, and predictive control under dynamic space conditions. \\
\hline

Digital Twin Network (DTN) \cite{yang_systematic_2021-1}\cite{tu_research_2023}\cite{zhou_hierarchical_2023}\cite{zhu_knowledge_2021}\cite{wei_data-driven_2021}\cite{chen_data_2023}\cite{wang_elastic_2023}\cite{duong_machine_2023}\cite{gao_plotinus_nodate}\cite{mao_digital_2024}\cite{lai_space_2024}
& Extension of DT concepts from individual assets to full communication networks for monitoring, diagnosis, simulation, and control. 
& Supports closed-loop network management, data-driven routing, network life-cycle support, safer testing, and improved operational visibility. 
& In satellite/NTN settings, major challenges remain in synchronization delay, model accuracy, scalability, limited onboard computing/storage, and real-time twinning of fast-moving LEO constellations. 
& Motivates a Digital Twin \emph{Satellite} Network that is designed specifically for satellite dynamics, asynchronous updates, resource limits, and predictive network adaptation. \\
\hline

SDN-based Satellite Network Operations \cite{chen_dijkstra-based_2023}\cite{zhang_deep_2024}\cite{wei_iris_2025}\cite{fu_reinforcement_2023}\cite{huang_hybrid_2025}\cite{sireesha_enhanced_2025}\cite{liang_green_2025}
& Separation of control plane and data plane for programmable and centralized satellite network management. 
& Improves flexibility, scalability, traffic engineering, routing control, and resource allocation in LEO/SAGIN environments. 
& Most approaches are reactive, depend on current-state snapshots, and focus mainly on network metrics while overlooking satellite bus conditions such as power, battery, and thermal state. 
& Motivates integrating network control with real-time physical satellite awareness so that routing and management become proactive and hardware-aware. \\
\hline

O-RAN-based Satellite/NTN Operations \cite{pasumarthy_demonstration_2024}\cite{firouzi_o2_2025}\cite{lee_3d_2025}\cite{baena_space-o-ran_2026}
& Open, disaggregated RAN architecture with RU/DU/CU split and RIC-based intelligent control for NTN. 
& Supports interoperability, flexible deployment, onboard/edge intelligence, xApps/rApps, and layered near-real-time/non-real-time control. 
& Existing studies often assume simplified onboard computing conditions and emphasize communication metrics while underrepresenting platform constraints such as power, thermal variation, and hardware degradation. 
& Motivates coupling O-RAN control with a high-fidelity satellite DT that continuously reflects the physical operating state of the satellite and orbit environment. \\
\hline

\textbf{Overall Insight}
& \multicolumn{4}{p{14.0cm}|}{Existing DT, DTN, SDN, and O-RAN studies each provide important building blocks for intelligent satellite networking, but none alone fully offers a real-time, bidirectional, predictive, and resilience-oriented framework that jointly considers network behavior and satellite physical state. This gap motivates the proposed Digital Twin Satellite Network (DTSN) in Section III.} \\
\hline

\end{tabular}
\end{table*}

\subsection{SDN-based Satellite Network Operations}
The paradigm shift in satellite networking has been driven by the adoption of SDN by separating the control plane from the data plane, SDN removes the tight coupling found in traditional architectures and allows the network to be managed in a more flexible and programmable way \cite{chen_dijkstra-based_2023}\cite{zhang_deep_2024}. In traditional satellite networks, each satellite must handle both complex routing decisions and packet forwarding at the same time. As constellations grow larger, this approach becomes impractical for satellites with limited computing resources to sustain efficient network operation \cite{wei_iris_2025}\cite{fu_reinforcement_2023}. By moving control functions, such as routing computation and network configuration from individual satellites to centralized entities (e.g., ground stations, network control centers, and geostationary satellites), satellite networks can be managed more efficiently while reducing the processing burden on resource-limited onboard systems \cite{huang_hybrid_2025}. This redistribution of network functions allows LEO satellites to focus mainly on data forwarding, making the overall system more flexible, scalable, and better suited to the growing demands of next-generation communication services \cite{sireesha_enhanced_2025}.

The centralized global perspective inherent in SDN architectures facilitates sophisticated dynamic operations within SAGIN and LEO constellations \cite{wei_iris_2025}. By separating network control from the underlying hardware, SDN controllers can better handle changing network topologies and intermittent links through real-time traffic management and automated resource allocation \cite{fu_reinforcement_2023}. Recent studies show that SDN can effectively balance network traffic by shifting data away from congested nodes and emerging bottlenecks, which helps reduce queuing delays and overall end-to-end latency \cite{chen_dijkstra-based_2023}. Furthermore, incorporating advanced data plane technologies such as P4-programmable switches and Segment Routing IPv6 (SRv6) enables more precise Quality of Service (QoS) control and supports traffic-engineered paths across heterogeneous, multi-domain network environments \cite{liang_green_2025}.

A critical evaluation of current SDN-based satellite networking studies reveals several limitations that affect their long-term reliability. In particular, most existing routing and management approaches are reactive, meaning that network adjustments are made only after link failures or congestion have already occurred. Many SDN controllers make decisions based only on the current network state or static snapshots, which often leads to suboptimal performance. This is because they do not account for the fast-changing topology of satellite constellations or the unpredictable behavior of inter-satellite links (ISLs) \cite{chen_dijkstra-based_2023}\cite{wei_iris_2025}. In addition, most existing SDN-based satellite studies pay little attention to the physical condition of the satellites bus. They mainly focus on network-level metrics such as bandwidth, delay, and packet loss, while overlooking important onboard factors like power availability, battery status, and thermal conditions. This separation between network control decisions and the physical state of the satellite platform suggests that SDN alone is not sufficient to ensure reliable and resilient satellite operation. Addressing this limitation requires a predictive DT framework that links network control with real-time hardware conditions and supports proactive resource management.

\subsection{O-RAN Satellite Network Operations}
The transition from traditional, tightly integrated satellite payloads to disaggregated architectures with open interfaces is a key step in the development of next-generation Non-Terrestrial Networks (NTNs) \cite{pasumarthy_demonstration_2024}. Traditionally, satellite communication systems relied on closed and tightly integrated control architectures, which offered limited programmability and made it difficult to scale or reconfigure networks dynamically. The O-RAN paradigm facilitates a significant shift by breaking the next-generation Node B (gNB) into separate, interoperable components, namely the Radio Unit (RU), Distributed Unit (DU), and Central Unit (CU). This separation allows components from different vendors to work together through standardized open interfaces, giving network operators the flexibility to choose the most suitable solutions for different parts of the SAGIN \cite{firouzi_o2_2025}. For example, placing only the RU or DU functions onboard satellites can simplify payload design and reduce launch costs, while enabling regenerative architectures in which satellites actively process signals instead of merely forwarding them as transparent relays \cite{lee_3d_2025}.

A central element that enables intelligent operation in this architecture is the RAN Intelligent Controller (RIC), which uses AI-based applications to support closed-loop network control and optimization. Recent literature highlights the importance of moving computing and control functions directly to the satellite edge in order to reduce the latency caused by long non-terrestrial communication links. \cite{baena_space-o-ran_2026}. Near-real-time xApps running on onboard or space-based RICs manage time-critical functions such as beam steering, interference control, and traffic scheduling, with control cycles typically ranging from 10 ms to 1 s. Simultaneously, non-real-time rApps operating within the terrestrial Service Management and Orchestration (SMO) layer handle longer-term functions such as global policy control and the retraining of AI models, often supported by high-fidelity DT simulations \cite{firouzi_o2_2025}\cite{lee_3d_2025}. This layered control structure, supported by the A1 and E2 interfaces, allows satellites to act as autonomous edge nodes. The A1 interface supports policy exchange and AI model updates between the SMO and the near-real-time RIC, while the E2 interface enables real-time monitoring and control of the underlying network elements \cite{firouzi_o2_2025}\cite{baena_space-o-ran_2026}.

However, a closer review of existing O-RAN–based NTN studies reveals an important gap related to the real operating conditions of small satellites. Most proposed frameworks focus on optimizing communication performance, such as data rates, compression efficiency, and handover latency, while often assuming ideal or highly simplified conditions for the onboard computing infrastructure. Many current approaches do not incorporate key physical constraints of the satellite platform, such as limited power availability, thermal variations, and hardware degradation caused by radiation, into their control frameworks. Instead, they mainly concentrate on link-level performance, without considering how communication demands directly affect the satellite’s energy capacity and thermal limits. For O-RAN to be deployed reliably in satellite systems, the network control layer needs to be tightly synchronized with a high-fidelity DT that continuously reflects the changing physical condition of the satellite and its operating environment in orbit.

Table~\ref{tab:related_work_summary} summarizes the main research streams reviewed in this section. Although DT, DTN, SDN, and O-RAN each provide important capabilities for intelligent satellite networking, none of these technologies alone offers an integrated framework that supports real-time digital twinning, predictive control, and hardware-aware resilience for satellite operations. Furthermore, most existing DTN studies for satellite systems mainly discuss architectural design, synchronization issues, or optimization concepts at a theoretical level. These gaps in existing approaches motivate the DTSN framework introduced in Section III. In contrast, this work presents a closed-loop DTSN framework and evaluates it through a cross-domain co-simulation over a large LEO constellation. Rather than focusing on isolated performance metrics, this approach examines how the digital twin responds to concurrent operational threats, including kinematic drift, spontaneous component failure, and adversarial jamming, in order to assess its resilience-oriented control capabilities.

\begin{figure*}[!t]
\centering
\includegraphics[width=\textwidth]{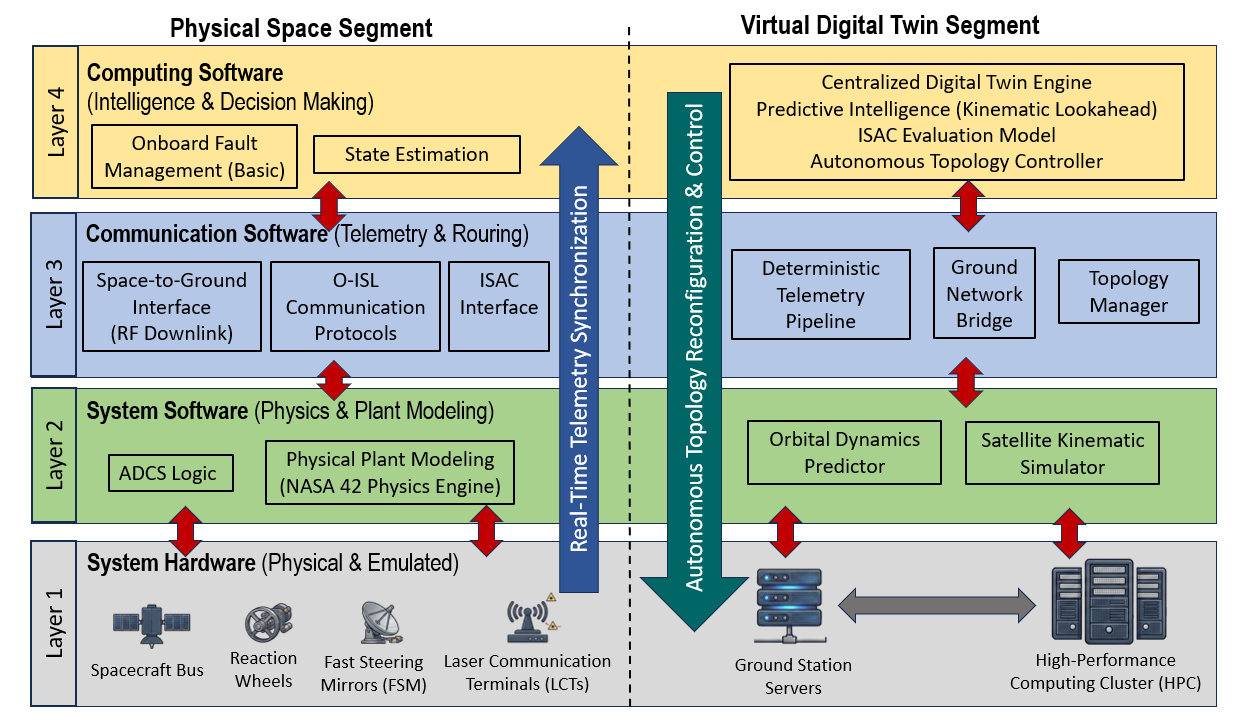}
\caption{The proposed Digital Twin Satellite Network (DTSN) layered architecture. The framework establishes a continuous, bidirectional closed-loop control system between the physical space segment and the virtual replica hosted on ground station servers. By decoupling the intensive computing software from the system hardware, predictive routing intelligence is executed without exceeding the strict Size, Weight, and Power (SWaP) constraints of LEO spacecraft.}
\label{fig:DTSN_design}
\end{figure*}

\section{Digital Twin Satellite Network (DTSN)}

Recent holistic operational architectures, such as SatAIOps \cite{hu_sataiops_2023}, have established that future satellite network management must integrate artificial intelligence across all mission phases. Within these advanced operational paradigms, digital twins are explicitly identified as a critical application domain capable of bridging ground-based AI services with real-time spacecraft telemetry. Addressing this need, the proposed DTSN scheme is designed to support the remote management of LEO satellite constellations, including large-scale mega-constellations. Unlike traditional static simulations, it operates as a real-time bidirectional control framework that continuously interacts with the physical network to maintain reliable operation in the space environment.

\subsection{DTSN Scheme Design}

To address the strict SWaP constraints of LEO satellites, the framework separates computationally intensive tasks from the physical space segment and offloads them to high-performance virtual replicas hosted on ground station servers.

As shown in Fig. \ref{fig:DTSN_design}, the DTSN architecture is organized across two domains: the Physical Space Segment and the Virtual DT Segment. It is further structured into four operational layers.

\begin{itemize}
    \item Layer 1 – System Hardware

    This layer includes the physical and emulated hardware infrastructure. In the space segment, it consists of the LEO satellite bus, reaction wheels, Fast Steering Mirrors (FSMs), and Laser Communication Terminals (LCTs). In the ground segment, it includes the ground station servers and High-Performance Computing (HPC) resources that support the DT operations.
    \item Layer 2 – System Software (Physics and Plant Modeling)

    This layer governs the laws of physics and baseline operating environments. On the physical satellite, this is the onboard Attitude Determination and Control System (ADCS) logic. In the virtual segment, it utilizes high-fidelity physical plant modeling (such as the NASA 42 Physics Engine) to continuously calculate orbital dynamics and spacecraft kinematics.

     \item Layer 3 – Communication Software (Telemetry and Routing)

     This layer functions as the data exchange bridge between the physical and virtual domains. It manages the space-to-ground RF telemetry link, O-ISL link communication protocols, and the ISAC interface within the constellation. On the ground side, it includes a deterministic telemetry pipeline and network bridge that support low-latency data ingestion and communication.

      \item Layer 4 – Computing Software (Intelligence and Decision Making)

      Because of onboard processing limitations, the physical satellite performs only lightweight state estimation and basic fault handling, e.g. failed component isolation, activating safe mode, etc. More complex computation is shifted to the ground-based centralized DT engine. This layer hosts the predictive intelligence modules of the DTSN, including kinematic lookahead prediction, ISAC-based link evaluation, hardware-health risk assessment, security anomaly interpretation, and the autonomous topology control function.

      The main capability of the DTSN scheme depends on real-time telemetry synchronization between the physical constellation and its DT. The twin continuously receives satellite attitude telemetry, including roll, pitch, and yaw measurements, through a deterministic data pipeline. With this synchronized state awareness, the centralized ground twin can evaluate network conditions in advance and send autonomous topology reconfiguration commands to the physical constellation. This predictive control process allows the network to respond proactively to multi-domain disturbances while avoiding heavy computational load on SWaP-constrained LEO satellites.

\end{itemize}

\begin{figure*}[!t]
\centering
\includegraphics[width=\textwidth]{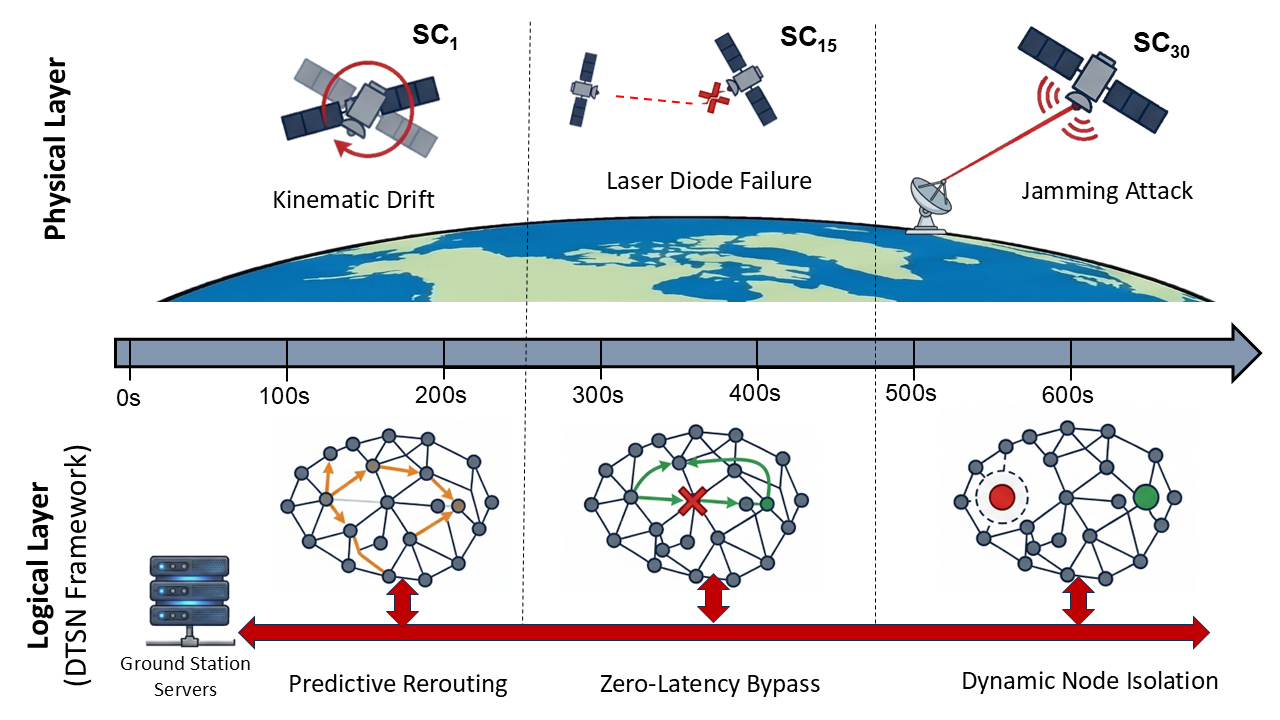}
\caption{Cross-domain co-simulation scenario used to validate the proposed Digital Twin Satellite Network (DTSN) framework. The upper layer illustrates the physical-domain disturbances introduced during the 600~s simulation window, including kinematic drift in SC$_{1}$, instantaneous laser diode failure in SC$_{15}$, and adversarial jamming of SC$_{30}$. The lower layer shows the corresponding logical DTSN responses, namely predictive rerouting, zero-latency bypass, and dynamic node isolation, executed by the ground-based digital twin to preserve resilient constellation operation.}
\label{fig:simulation}
\end{figure*}

\subsection{Core Operational Features of the DTSN}
\begin{enumerate}
    \item Integrated Sensing and Communication (ISAC)
    
    DTSN utilizes ISAC by combining communication and sensing functions as unified framework where data transmission and environmental sensing operate together \cite{kim_survey_2022}. ISAC allows both functions to share the same spectrum, hardware platform, and signal processing structure. This integrated approach improves spectral efficiency and reduces the interference that usually occurs when separate sensing and communication systems operate in parallel \cite{sam_learning-based_2025}.

    To implement ISAC under the strict SWaP constraints of small LEO satellites, the proposed DTSN may be realized through a dual-functional optical payload design in which the same O-ISL hardware supports both communication and state sensing. This design choice is one practical realization of the DTSN framework, not a prerequisite for its operation. In general, DTSN can also work with separate sensing and communication payloads, provided that their outputs are synchronized in the digital twin. In the present study, we focus only on an optical-link realization; therefore, the sensing functionality is derived from O-ISL observables rather than from an integrated radar payload. In this case, the digital coherent optical transceivers support high-speed data transmission \cite{fuchs_optical_2023} while also extracting optical-link telemetry from the received optical signal. By continuously monitoring signal intensity variations, state of polarization (SOP) changes, and optical phase fluctuations, the LCT can assess channel conditions and detect physical anomalies, such as thermal effects caused by solar irradiance, directly at the hardware level.

    The handling of the large volume of telemetry generated by these dual-functional optical links requires efficient onboard processing \cite{ma_integrated_2024}. To avoid overloading the space-to-ground downlink, the satellite bus incorporates mobile Edge Computing (MEC) units that analyze raw signal features locally instead of transmitting continuous raw data streams \cite{wen_survey_2025}. This edge-based ISAC approach allows the remote DT to maintain high-fidelity synchronization with the physical satellite bus while preserving energy and bandwidth efficiency across the constellation.

    In the present study, ISAC is demonstrated through the continuous mapping of spacecraft attitude behavior to O-ISL link quality within a cross-domain co-simulation. In particular, the kinematic disturbance scenario shows how physical motion can be translated into communication degradation in real-time. More advanced dual-functional optical sensing features, such as phase-aware and polarization-aware diagnostics, remain part of the broader DTSN design perspective for future implementation.

    \item Quantum Sensing for Fault Diagnosis

The second feature is quantum sensing for future fault diagnosis. While ISAC and classical telemetry are effective for detecting performance degradation, they may not always be sufficient to isolate the precise physical cause of a fault \cite{gizzi_autonomous_nodate}. In practical satellite systems, similar communication symptoms may originate from different mechanisms, such as fine attitude jitter, thermal deformation, timing instability, structural vibration, or subtle payload misalignment \cite{liu_impact_2025}. Distinguishing between these causes is important for robust autonomous control.

To improve diagnostic accuracy, we consider quantum sensing as a possible future extension of layer 3 of DTSN architecture in Fig. \ref{fig:DTSN_design}. Quantum sensing technologies are known for their very high sensitivity to small physical changes \cite{kantsepolsky_exploring_2023}. In future satellite systems, this capability could enable more precise measurements of parameters such as motion, timing variations, field disturbances, and structural anomalies compared to conventional sensing methods \cite{kundu_quantum_2023}.

It is important to clarify that quantum sensing is considered in this work as a forward-looking architectural extension rather than as a fully mature flight-ready sensing subsystem. In the present case study, its role is used to motivate how future near-zero-drift sensing inputs could strengthen early anomaly discrimination within the DTSN. The current co-simulation still relies primarily on synchronized telemetry and digital twin intelligence for operational control, while preserving a framework that can incorporate more advanced sensing modalities as they become practical for LEO deployment.

Thus, while ISAC supports broad real-time observability, quantum sensing can be viewed as a future precision diagnostic layer that strengthens the ability of the DTSN to move from fault detection toward fault discrimination.

    \item Proactive Intelligence
    
   The third core feature of the DTSN is proactive intelligence, which represents the decision-making capability of the DT. In the proposed framework, intelligence refers specifically to the capability of the twin to anticipate system evolution, evaluate alternative control responses in the virtual domain, and select a suitable operational action before the physical network experiences service degradation.

   This feature is essential because satellite networks are highly dynamic, and network reliability is challenged by frequent link handovers, topological fluctuations, and uncertain hardware failures \cite{wei_iris_2025}\cite{fu_reinforcement_2023}. A reactive controller that waits until a link is already broken may incur packet loss, rerouting delay, or wider service disruption. In contrast, the proactive intelligence of the DTSN uses the synchronized twin state to estimate short-term future conditions and determine whether the active communication path is likely to remain reliable.

   In the present paper, this concept is implemented in Section IV through a cross-domain decision framework that evaluates physical, hardware, and security disturbances within the same synchronized timeline. For kinematic degradation, the twin forecasts the near-future effect of attitude drift on optical link alignment. For hardware anomalies, it uses historical and operational state trends to anticipate high-risk component failure. For adversarial interference, it identifies the onset of abnormal SNR behavior and supports resilient routing adaptation. In this way, proactive intelligence allows the DT to move beyond passive monitoring and take an active role in network operation.

  Although the present co-simulation includes a lightweight kinematic prediction model for the physical disturbance scenario, the DTSN framework is not limited to a single algorithmic method or a single disturbance class. More advanced implementations may incorporate data-driven forecasting, learning-based anomaly prediction, optimization-based control selection, or hybrid model-driven and AI-driven decision engines across physical, hardware, and security domains. The present framework is therefore general enough to support multiple intelligence mechanisms while preserving the same closed-loop operational principle.

    \item Dynamic Resilience
    
The fourth feature of the DTSN is dynamic resilience, which is the ultimate operational objective of the framework. In this work, resilience refers to the ability of the satellite network to maintain service continuity under disturbances through early awareness, predictive decision making, and timely control adaptation.

Satellite communication systems are exposed to many forms of uncertainty, including orbital dynamics, variable link geometry, platform disturbances, hardware limitations, and environmental effects \cite{baltaci_investigation_2023}. As a result, maintaining reliable service requires more than static routing or periodic reconfiguration. It requires a control framework that can continuously adapt to changing risk conditions in real-time. This is the role of dynamic resilience in the proposed DTSN.

Dynamic resilience results from the combined action of the previous three features. ISAC provides the operational awareness needed to detect emerging degradation. Proactive intelligence predicts how that degradation may evolve. The control layer then uses this information to preserve network continuity by selecting a more reliable path or operational state. In this sense, resilience is not treated as a separate function, but as the system-level outcome of the DTSN closed loop.

In the cross-domain satellite network scenario illustrated in Fig. \ref{fig:simulation}, resilience is demonstrated across three different forms of disruption. When a gradual kinematic drift develops in SC$_{1}$, the DT forecasts the resulting communication degradation and supports predictive rerouting before the link reaches a critical state. When SC$_{15}$ experiences an instantaneous hardware failure, the framework bypasses the affected node through a precomputed alternative path. When SC$_{30}$ is subjected to adversarial jamming, the DTSN isolates the affected node, preserves wider network continuity, and reintegrates the node after recovery. These behaviors provide concrete examples of resilience-oriented network management in a satellite setting.

Therefore, dynamic resilience is a defining capability of the proposed DTSN, since it preserves communication service through predictive twin-driven adaptation under changing physical network conditions.

\end{enumerate}

\subsection{Discussion and Relation to the Case Study}

The proposed DTSN framework provides the conceptual and operational basis for the experimental validation presented in Section IV. In particular, the case study is intentionally designed to demonstrate the interaction among the four core features introduced above.

First, the simulation uses real-time spacecraft telemetry as the synchronization input of the DT, reflecting the sensing and state update functions of the DTSN. Second, the ISAC model links physical attitude behavior to communication quality, providing an operational example of sensing-communication integration within the framework. Third, the predictive decision logic extends beyond kinematic estimation to include proactive assessment of hardware degradation risk and security-related link disruption. Finally, the control response illustrates dynamic resilience through predictive rerouting, zero-latency bypass, and temporary node isolation, depending on the disturbance type observed in the constellation.

Accordingly, Section IV should be read as a constellation-scale cross-domain validation of the DTSN scheme. The objective is to show that a satellite-specific DT can combine real-time physical awareness, multi-domain fault interpretation, and predictive control to improve operational reliability and resilience under realistic LEO network conditions.

\section{Case Study and Experimental Results}

To validate the proposed DTSN framework, this section presents a comprehensive cross-domain co-simulation, designed to demonstrate the capability of the framework to support ISAC, predictive intelligence, and autonomous network resilience in a mega-constellation setting. The case study examines the framework under three disturbance domains, namely physical, hardware, and security disturbances, which are injected simultaneously into a high-fidelity orbital environment over a continuous 600 s flight window.

\subsection{Simulation Architecture and Baseline Constellation Topology}

Before presenting the cross-domain threat emulation, the baseline physical topology of the space segment should be defined. The simulation is built on a LEO constellation configured according to a symmetric Walker Delta \(T/P/F\) geometry, specifically a \(60/6/3\) topology.

As illustrated in Table \ref{tab:orbital_params}, the total number of spacecraft \((T = 60)\) is evenly distributed across \(P = 6\) circular orbital planes, resulting in 10 satellites per plane. To provide uniform global coverage, the orbital planes are symmetrically arranged around the Earth with a Right Ascension of the Ascending Node (RAAN) separation of \(60^\circ\) \((\Delta \Omega = 360^\circ / P)\). Within each orbital plane, the satellites maintain a fixed intra-plane true anomaly separation of \(36^\circ\) \((\Delta \nu = 360^\circ / 10)\).

A relative phasing parameter of \(F = 3\) is further applied to the constellation. In the \(60/6/3\) configuration, satellites in the adjacent eastern plane are shifted by a phase angle of \(18^\circ\) \((\Delta \Phi = F \times 360^\circ / T)\) relative to the equatorial crossing nodes of the current plane. This \(18^\circ\) offset helps prevent collisions at orbital intersections while also establishing the geometric line-of-sight conditions required to support continuous optical inter-satellite links (O-ISLs).

The physical dynamics of the constellation are simulated using the NASA 42 Spacecraft Simulator. The physics engine is configured to generate a constellation-wide master telemetry matrix at a frequency of 10 Hz, containing eleven physical parameters for all 60 spacecraft, including roll, pitch, yaw, and three-dimensional velocity vectors. A custom Python-based DT bridge continuously ingests this synchronized dataset and overlays logical network and security faults onto the physical timeline in order to evaluate the multi-layer response mechanisms of the DTSN.

\begin{table}[htbp]
\caption{Orbital Parameters of the Baseline Constellation}
\label{tab:orbital_params}
\centering
\begin{tabular}{|l|c|}
\hline
\textbf{Parameter} & \textbf{Value} \\
\hline
\hline
Constellation Geometry & Walker Delta \\
\hline
Topology Configuration ($T/P/F$) & $60/6/3$ \\
\hline
Total Satellites ($T$) & 60 \\
\hline
Number of Orbital Planes ($P$) & 6 \\
\hline
Satellites per Plane & 10 \\
\hline
Phasing Parameter ($F$) & 3 \\
\hline
Orbital Inclination & $55^\circ$ \\
\hline
RAAN Spacing ($\Delta\Omega$) & $60^\circ$ \\
\hline
Intra-Plane True Anomaly Spacing ($\Delta\nu$) & $36^\circ$ \\
\hline
Inter-Plane Phase Shift ($\Delta\Phi$) & $18^\circ$ \\
\hline
\end{tabular}
\end{table}

\subsection{Cross-Domain Threat Emulation and DTSN Response}

\begin{figure}[htbp]
    \centering
    \includegraphics[width=\columnwidth]{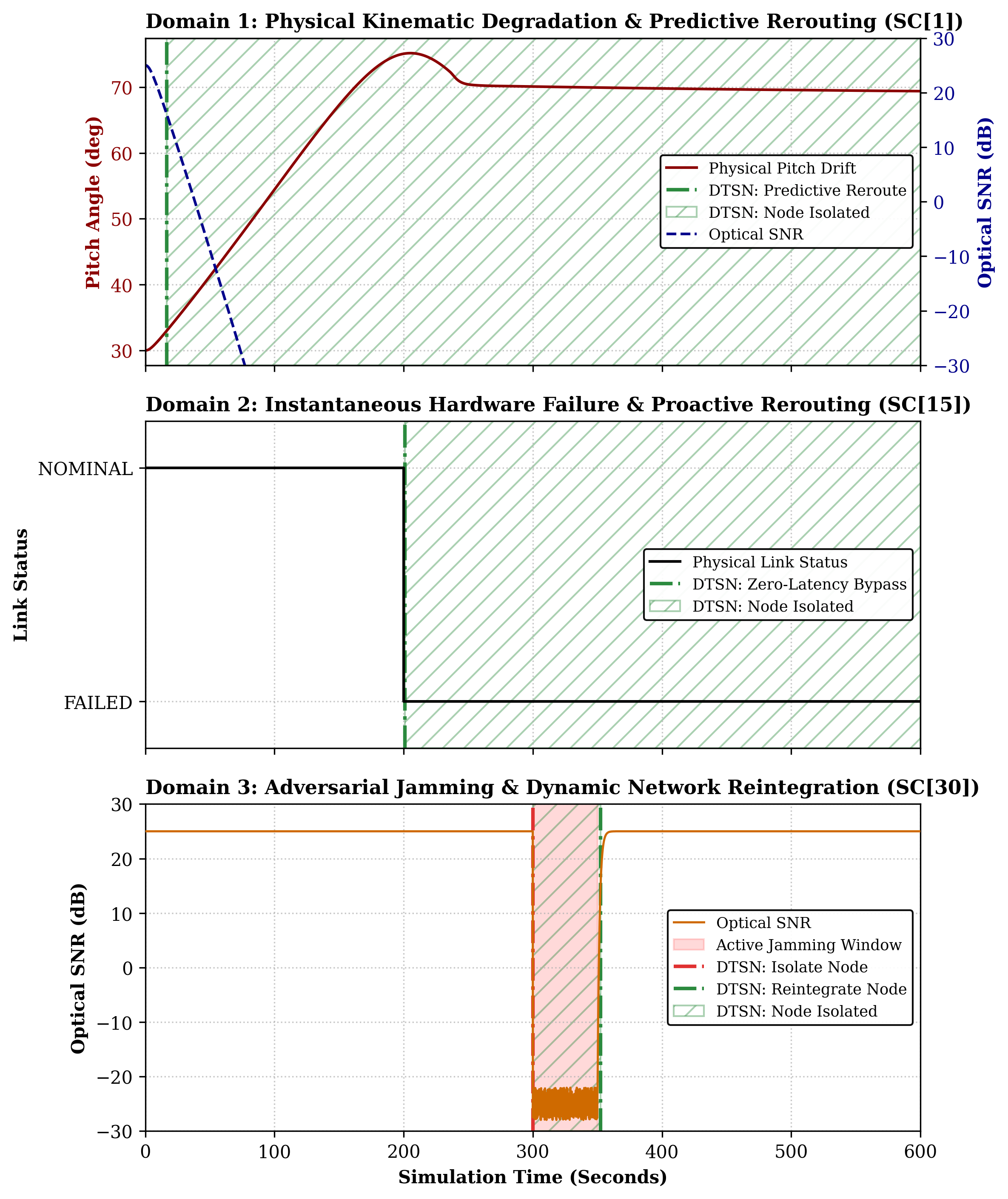}
    \caption{Cross-domain Digital Twin Satellite Network (DTSN) response during the 600~s co-simulation. The top panel shows the gradual kinematic degradation of SC$_{1}$, where the digital twin maps pitch drift to optical SNR degradation and triggers predictive rerouting before the critical threshold is reached. The middle panel shows the instantaneous hardware failure of SC$_{15}$, which is mitigated through a proactive zero-latency bypass. The bottom panel shows the adversarial jamming of SC$_{30}$, including the resulting SNR collapse, temporary node isolation, and safe reintegration after receiver recovery.}
    \label{fig:tri_fault_routing}
\end{figure}

To evaluate the DTSN architecture without introducing the high computational cost of computing network layer states directly within a continuous Newtonian physics engine, this study adopts a synthetic fault injection methodology based on a cyber-physical overlay. In this setting, the continuous physical space environment, simulated natively in NASA 42, is intentionally decoupled from the discrete logical network environment managed by the Python-based digital twin. Physical kinematic anomalies are solved dynamically within the orbital mechanics engine, whereas binary hardware failures and adversarial cyber attacks are injected synthetically as discrete logical overrides on top of the synchronized baseline telemetry. This cross-domain co-simulation approach provides a practical and scalable way to evaluate multi-vector network threats while preserving a mathematically consistent and collision-free orbital foundation, the simulation parameters are summarized in Table \ref{tab:sim_params}. The complete synchronized timeline of this cross-domain emulation, detailing the physical anomalies, hardware failures, cyber-physical attacks, and the corresponding autonomous routing decisions executed by the DTSN, is visualized in Fig. \ref{fig:tri_fault_routing}.

\begin{table}[htbp]
\caption{Co-Simulation and Cross-Domain Threat Parameters}
\label{tab:sim_params}
\centering
\begin{tabular}{|l|c|}
\hline
\textbf{Parameter} & \textbf{Value} \\
\hline
\hline
Physics Engine & NASA 42 Simulator \\
\hline
Total Simulation Duration & 600 s \\
\hline
Telemetry Extraction Rate & 10 Hz \\
\hline
Nominal Optical SNR & 25.0 dB \\
\hline
Minimum SNR Operational Threshold & 15.0 dB \\
\hline
Scenario A: Kinematic Drift (SC$_{1}$) & $T=0$ s \\
\hline
Scenario B: Hardware Burnout (SC$_{15}$) & $T=200$ s \\
\hline
Scenario C: Jamming Attack (SC$_{30}$) & $T=300$ s to $350$ s \\
\hline
Sensor Recovery Time Constant ($\tau$) & 1.5 s \\
\hline
\end{tabular}
\end{table}

    Scenario A: Kinematic Degradation via ISAC and Quantum Sensing

    In this scenario, quantum sensing is represented as a forward-looking sensing input within the digital twin architecture rather than as a fully simulated onboard flight subsystem. In highly directional optical communications, link quality is strictly dependent on the physical pointing accuracy maintained by the Fast Steering Mirrors (FSM) \cite{xue2025high}. To integrate physical sensing with communication performance, a core paradigm of ISAC \cite{liu2022integrated}, the DT dynamically maps the satellite's angular velocity to SNR degradation. This mathematical mapping illustrates the practical application of the ISAC concept within our framework. Since physical pointing accuracy and communication quality are inherently linked in O-ISL links, the DT directly translates a physical movement, specifically, the angular velocity of the FSM, into a communication metric (SNR degradation). Through this direct translation, sensing a mechanical disturbance actively functions as a real-time channel estimation tool, illustrating the dual sensing-communication relationship expected in ISAC systems.
    Beginning at \(T = 0\) s, Spacecraft 1 (SC$_{1}$) experiences gradual momentum wheel desaturation, which leads to a continuous pitch drift. The DTSN addresses this mechanical deviation through the combined use of quantum sensing and ISAC. In the proposed architecture, next-generation quantum-enhanced gyroscopes in the system hardware layer (layer 1) of the DTSN are assumed to detect micro-radian attitude anomalies well before they develop into link failure. Although chip-scale Cold Atom Interferometers (CAIs) are currently at Technology Readiness Level (TRL) 5--6 and still face important miniaturization challenges related to (SWaP) \cite{abend_technology_2023}, their inclusion in the framework is intended as a forward-looking design direction. In particular, the DTSN is structured to ingest this near-zero-drift kinematic telemetry once such sensors become practically deployable in LEO systems.
    Because Free-Space Optical (FSO) lasers exhibit a Gaussian intensity profile, the pointing loss in decibels is proportional to the square of the pointing error \cite{farid2007outage}. Assuming the FSM tracking lag is directly proportional to the physical angular velocity of the spacecraft during an attitude disturbance, the DT evaluates the real-time SNR utilizing a quadratic heuristic model \cite{kaushal2017optical} as shown in (\ref{eq:quad_hmodel}):

\begin{equation}
SNR_{ISL} = SNR_{nominal} - \alpha \cdot (Pitch\_Rate)^2
\label{eq:quad_hmodel}
\end{equation}

where \(SNR_{\mathrm{nominal}}\) represents the baseline optical link quality of \(25.0\,\mathrm{dB}\), and \(\alpha\) denotes the hardware degradation coefficient representing the optical beam divergence and FSM tracking latency. To maintain reliable data transmission, the network enforces a strict minimum threshold of $SNR_{min} = 15.0$ dB, allowing for a maximum link margin degradation of 10.0 dB. Consequently, the coefficient was empirically calibrated to $\alpha = 127.5$. This mathematically maps the FSM's critical tracking failure boundary ($0.28^\circ$/s) to the exact 10.0 dB allowable pointing loss, cleanly separating natural orbital perturbations from severe attitude anomalies.

Traditional network routing in dynamic space environments usually relies on reactive fault detection mechanisms. Such approaches may introduce latency, increased control overhead, and packet loss during the network topology reconfiguration process \cite{zhao2024adaptive}. To address this limitation, recent studies have explored the use of DTs to enable proactive switching strategies and congestion prediction for inter-satellite links (ISLs) \cite{li2023research}. 

Following this idea, the proposed DTSN framework adopts a kinematic predictive intelligence model that anticipates optical tracking mirror failures before the physical communication link is disrupted. This predictive model implements the “Proactive Intelligence” feature introduced in Section III-B3.

The DT estimates the instantaneous angular velocity of the satellite and predicts the tracking deviation within a lookahead window ($T_{\text{lookahead}} = 5.0$ s) as shown in (\ref{eq:predicted_deviation}):

\begin{equation}
\theta_{\mathrm{predicted\_deviation}} = \left| \mathrm{Pitch\ Rate} \right| \times T_{\mathrm{lookahead}}
\label{eq:predicted_deviation}
\end{equation}
By continuously relating the physical deviation to the declining optical SNR, the DTSN can identify the root cause as mechanical misalignment. This allows the system to perform a preemptive network topology reroute before the SNR falls below the minimum operational threshold of \(15.0\,\mathrm{dB}\).

   Scenario B: Hardware Failure and Proactive Intelligence

   In contrast to the gradual nature of kinematic drift, hardware anomalies often appear as sudden binary failures. At \(T = 200\) s, Spacecraft 15 (SC$_{15}$) experiences a spontaneous optical laser diode burnout, which causes the link status to drop immediately from nominal operation to failure.

   Because this type of hardware failure does not exhibit a clear kinematic precursor, a conventional telemetry system would only observe an abrupt and otherwise unexplained network outage. The DTSN addresses this limitation through its proactive intelligence capability. By continuously applying predictive machine learning models to the historical power consumption patterns, thermal profiles, and component lifetime data of the constellation, the digital twin identifies a high probability of diode failure before the physical event occurs. In the present co-simulation, these health indicators are represented as synthetic logical-layer inputs aligned with the synchronized physical telemetry, allowing the DT to emulate hardware-risk assessment without embedding full component-aging physics in the orbital model. As a result, when the burnout takes place, the proactive routing tables have already determined an alternative optical path, allowing the system to bypass SC$_{15}$ immediately, avoid data bottlenecks, and maintain traffic rerouting without additional delay.

    Scenario C: Adversarial Jamming and Dynamic Resilience
    
    The final scenario evaluates the security domain of the proposed architecture. Between \(T = 300\) s and \(T = 350\) s, Spacecraft 30 (SC$_{30}$) passes through a high-risk orbital region and is exposed to an external cyber-physical attack in the form of a terrestrial laser dazzler. This adversarial jamming event forces the optical Signal-to-Noise Ratio (SNR) to fall sharply from its nominal value of \(25\) dB to a chaotic noise floor of approximately \(-25\) dB, thereby triggering a critical jamming alarm.
    
    After the attack ends at \(T = 350\) s, the optical receiver enters a recovery phase characterized by thermal cooling and focus realignment. To represent this hardware recovery process within the digital twin, an exponential sensor recovery model is introduced \cite{majumdar_advanced_2015}:
    \begin{equation}
    SNR_{\mathrm{recovery}}(t) = SNR_{\mathrm{nominal}} - \beta \, e^{-\frac{(t - t_{\mathrm{end}})}{\tau}}
    \label{eq:snr_recovery}
    \end{equation}
    where \(t_{\mathrm{end}}\) denotes the end time of the jamming attack, \(\beta\) represents the maximum degradation caused by the jammer (e.g., \(50.0\) dB), and \(\tau\) is the hardware-specific sensor recovery time constant which is designated as 1.5 s. This value serves as an aggregate heuristic parameter representing the combined delays of two sequential physical recovery phases: the thermal relaxation of the saturated Avalanche Photodiode (APD) substrate in a vacuum environment, and the subsequent phase re-acquisition timeline required by the receiver's Phase-Locked Loop (PLL) to re-synchronize with the optical carrier frequency \cite{gardner_phaselock_2005}.
    
    This highly dynamic and hostile condition activates the Dynamic Resilience protocol of the DTSN. Once the digital twin detects the chaotic SNR fluctuations associated with adversarial jamming, it immediately isolates the affected node in order to prevent corrupted data from propagating through the ISLs. The resilient control architecture then autonomously shifts the transmission frequencies or optical wavelengths of neighboring satellites to bypass the jammed spectrum. After SC$_{30}$ exits the attack interval and its optical receiver recovers above the safety threshold of \(15\) dB according to (\ref{eq:snr_recovery}), the DTSN reintegrates the node into the primary routing matrix without disrupting the continuity of network operation.

   \subsection{Co-Simulation Outcomes and Synthesis.}

   By jointly tracking a gradual kinematic fault, an instantaneous hardware failure, and a chaotic external security attack, the closed-loop co-simulation produces a synchronized dataset containing 6{,}000 frames. Within this unified observation window, the digital twin is able to isolate and classify the overlapping network disruptions in real-time. These results indicate that the integration of high-fidelity orbital mechanics with a localized digital twin can provide the predictive intelligence and autonomous resilience needed to manage the strict SWaP constraints and rapidly changing topologies of next-generation LEO mega-constellations.


\section{Perspectives on Future Directions}

In this paper, we introduced the DTSN framework as a closed-loop architecture for reliable and intelligent operation of LEO constellations. By synchronizing physical spacecraft telemetry with ground-based virtual replicas, the proposed framework supports proactive network management. The case study showed that the DTSN can combine physical awareness, ISAC-based link evaluation, predictive intelligence, and resilience-oriented control within a cross-domain co-simulation that includes kinematic degradation, hardware failure, and adversarial interference. These results highlight the potential of DTSN for large-scale and adaptive satellite network operation.

Several future directions can further extend this framework. First, although the current ISAC implementation can detect performance degradation, it may not always identify the exact physical source of complex faults. Future work should therefore investigate the integration of quantum sensing into the DTSN architecture. Due to their high sensitivity to small physical changes, quantum sensors may improve the measurement of structural anomalies, timing variations, and field disturbances, thereby strengthening fault diagnosis and root-cause identification.

Second, the predictive intelligence engine can be further improved. Although the present study evaluates disturbances in multiple domains, the predictive mechanisms remain relatively lightweight and scenario-specific. Future versions may incorporate richer data-driven forecasting, learning-based anomaly prediction, and hybrid AI-assisted decision mechanisms to support more accurate, adaptive, and scalable control across large satellite constellations.

Finally, extending DTSN to large-scale satellite mega-constellations will require closer integration with SDN and O-RAN frameworks. Combining layered network control with a high-fidelity DT may support more scalable, resilient, and hardware-aware satellite network operation under the constraints of the space environment.


%


\section*{Acknowledgment}
We acknowledge the support provided by the Government of Canada and Natural Sciences and Engineering Research Council of Canada (NSERC), [funding reference number RGPIN-2022-03364], Research Manitoba, and the Digital Research Alliance of Canada.

\ifCLASSOPTIONcaptionsoff
  \newpage
\fi




\bibliographystyle{IEEEtran}
\bibliography{references}
\end{document}